\documentclass[twocolumn,prb,superscriptaddress,showpacs,amsmath,amssymb]{revtex4}
\usepackage{bm,graphicx,dcolumn}
\newcommand{\pdt}{\partial_{t}}

\begin{document}
\title{Ambegaokar-Baratoff relations of Josephson critical current in 
heterojunctions with multi-gap superconductors}

\affiliation{
CCSE, Japan Atomic Energy Agency, 
6-9-3 Higashi-Ueno Taito-ku, Tokyo 110-0015, Japan}
\affiliation{
Department of Physics, Keio University, 
3-14-1 Hiyoshi, Kohoku-ku, Yokohama 223-8522, Japan}
\affiliation{
Institute for Materials Research, Tohoku University, 2-1-1 Katahira
Aoba-ku, Sendai 980-8577, Japan} 
\affiliation{
CREST(JST), 4-1-8 Honcho, Kawaguchi, Saitama 332-0012, Japan}
\affiliation{
JST, TRIP, 5 Sambancho Chiyoda-ku, Tokyo 102-0075, Japan}
\author{Yukihiro Ota}
\affiliation{
CCSE, Japan Atomic Energy Agency, 
6-9-3 Higashi-Ueno Taito-ku, Tokyo 110-0015, Japan}
\affiliation{
CREST(JST), 4-1-8 Honcho, Kawaguchi, Saitama 332-0012, Japan}
\author{Noriyuki Nakai}
\affiliation{
CCSE, Japan Atomic Energy Agency, 
6-9-3 Higashi-Ueno Taito-ku, Tokyo 110-0015, Japan}
\affiliation{
CREST(JST), 4-1-8 Honcho, Kawaguchi, Saitama 332-0012, Japan}
\author{Hiroki Nakamura}
\affiliation{
CCSE, Japan Atomic Energy Agency, 
6-9-3 Higashi-Ueno Taito-ku, Tokyo 110-0015, Japan}
\affiliation{
CREST(JST), 4-1-8 Honcho, Kawaguchi, Saitama 332-0012, Japan}
\affiliation{
JST, TRIP, 5 Sambancho Chiyoda-ku, Tokyo 102-0075, Japan}
\author{Masahiko Machida}
\affiliation{
CCSE, Japan Atomic Energy Agency, 
6-9-3 Higashi-Ueno Taito-ku, Tokyo 110-0015, Japan}
\affiliation{
CREST(JST), 4-1-8 Honcho, Kawaguchi, Saitama 332-0012, Japan}
\affiliation{
JST, TRIP, 5 Sambancho Chiyoda-ku, Tokyo 102-0075, Japan}
\author{Daisuke Inotani}
\affiliation{
Department of Physics, Keio University, 
3-14-1 Hiyoshi, Kohoku-ku, Yokohama 223-8522, Japan}
\author{Yoji Ohashi}
\affiliation{
Department of Physics, Keio University, 
3-14-1 Hiyoshi, Kohoku-ku, Yokohama 223-8522, Japan}
\affiliation{
CREST(JST), 4-1-8 Honcho, Kawaguchi, Saitama 332-0012, Japan}
\author{Tomio Koyama}
\affiliation{
Institute for Materials Research, Tohoku University, 
2-1-1 Katahira Aoba-ku, Sendai 980-8577, Japan}
\affiliation{
CREST(JST), 4-1-8 Honcho, Kawaguchi, Saitama 332-0012, Japan}
\author{Hideki Matsumoto}
\affiliation{
Institute for Materials Research, Tohoku University, 
2-1-1 Katahira Aoba-ku, Sendai 980-8577, Japan}
\affiliation{
CREST(JST), 4-1-8 Honcho, Kawaguchi, Saitama 332-0012, Japan}
\date{\today}

\begin{abstract}
An extension of the Ambegaokar-Baratoff relation to a
 superconductor-insulator-superconductor (SIS) Josephson
 junction with multiple tunneling channels is derived. 
Appling the resultant relation to a SIS Josephson junction
 formed by an iron-based (five-band) and a single-band Bardeen-Cooper-Schrieffer
 (BCS) type superconductors, a theoretical bound of the Josephson
 critical current ($I_{\rm c}$) multiplied by the resistance of the
 junction ($R_{\rm n}$) is given. 
We reveal that such a bound is useful for identifying the pairing
 symmetry of iron-pnictide superconductors. 
One finds that if a measured value of $I_{\rm c}R_{\rm n}$
 is smaller than the bound then the symmetry is $\pm s$-wave, and
 otherwise $s$-wave without any sign changes.  
In addition, we stress that temperature dependence of 
$I_{\rm c}R_{\rm n}$ is sensitive to the difference of the gap functions
 from the BCS type gap formula in the above heterojunction. 
\end{abstract}

\maketitle

\section{Introduction}
Since the discovery of iron-based superconductors\,\cite{Kamihara;Hosono:2008,Rotter;Johrendt:2008,Sasmal;Chu:2008,Tapp;Guloy:2008,ZhiAn;ZhongXian:2008,Wang;Jin:2008,Hsu;Wu:2008},
their pairing symmetry has been intensively debated. 
According to the spin fluctuation mechanism associated with the Fermi
surface nesting, $\pm s$-wave
symmetry was proposed as a pairing
scenario\,\cite{Mazin;Du:2008,Kuroki;Aoki:2008,Graser;Scalapino:2009,Wang;Lee:2009,Yu;Li:2009,Cvetkovic;Tesanovic:2009,Kuroki;Aoki:2009}. 
However, the debate has not been settled down. 
The $\pm s$-wave symmetry is expected to be fragile
against non-magnetic impurities\,\cite{Onari;Kontani:2009}. 
Some experiment\,\cite{Guo;TanakayaMuromachi:2009} supported this idea,
while the
others\,\cite{Lee;Sato:2009,Sato;Miura:2010,Li;Xu:2009,Cheng;Wen:2010,Li;Zhang:2010}
presented controversial results. 
Hence, a direct and unambiguous evidence like the phase sensitive measurement 
in High-$T_{\rm c}$ cuprate superconductors\,\cite{Tsuei;Kirtley:2000} 
is now in great demand. 
In fact, a large number of the methods to seek a definite signature have been
examined, e.g., tunneling 
spectroscopy\,\cite{Onari;Tanaka:2009,Linder;Sudbo:2008,Sperstad;Sudbo:2009,Nagai;Hayashi:2009,Golubov;Tanaka:2009},
corner junctions\,\cite{Tsai;Hu:2009,Parker;Mazin:2009,Wu;Phillips:2009,Chen;Zhang:2009}, 
observation of half-integer flux-quantum jump\,\cite{Chen;Zhao;2009}, 
scanning tunnel microscopy\,\cite{Hanaguri;Takagi:2010}, and so on. 

Josephson junctions are sensitive devices reflecting superconducting
states of each electrode. 
Very recently, various types of Josephson junctions with iron-pnictide
superconductors were successfully fabricated and typical Josephson
effects were confirmed\,\cite{Zhang;Takeuchi:2009,Zhang;Saha;Takeuchi:2009,Kashiwaya;Kashiwaya:2009,Katase;Hosono:2010}. 
Among them, a Josephson junction between an iron-based  
and a conventional $s$-wave single-gap superconductors has
been regarded as a possible candidate to directly detect the pairing
symmetry of iron-based superconductors. 
The heterojunction system is theoretically
described by multiple tunneling channels, some of which are $\pi$ channels
and the others are $0$ ones\,\cite{Linder;Sudbo:2009,Ota;Matsumoto:2009}. 
Authors suggested anomalous critical current
reduction\,\cite{Ota;Matsumoto:2009}, Riedel
anomaly cancellation\,\cite{Inotani;Ohashi:2009}, and enlargement of the Josephson
vortex core\,\cite{Ota;Machida;Matsumoto:2010}. 

In this paper, we derive Ambagaokar-Baratoff
relation\,\cite{Ambegaokar;Baratoff:1963} in 
the heterojunctions with multiple tunneling channels and clarify that
a theoretical bound of $I_{\rm c}R_{\rm n}$ products distinguishes 
$\pm s$-wave from $s$-wave without any sign changes, which 
is simply denoted as $s$-wave throughout this paper. 
We examine two kinds of materials, 
$(\mbox{Ba,K})\mbox{Fe}_{2}\mbox{As}_{2}$ (122 compound) and
$\mbox{LaFeAs}(\mbox{O,F})$ (1111 compound) as the iron-based
superconducting electrode in the heterojunction. 
Employing the density of states (DOS) ratios and the superconduting gap
ratios given by five-band quasi-classical theory with the first-principles 
calculation\,\cite{Nakai;Machida:2009}, the theoretical bounds are
evaluated. 
The temperature dependences of $I_{\rm c}R_{\rm n}$ are also
demonstrated in both $\pm s$-wave and $s$-wave. 

The paper is organized as follows. 
Section \ref{sec:jc} is the derivation of the Ambegaokar-Baratoff
relation in the junction with multiple tunneling channels. 
Based on the result, we propose a criterion for 
identifying the pairing symmetry of iron-based superconductors. 
The key criterion is an upper bound of the Josephson critical
current for the $\pm s$-wave, which corresponds to a lower bound for the
$s$-wave.  
In Sec.\,\ref{sec:iron}, we apply this criterion to typical
iron-pnictide superconducting materials and theoretically confirm its
effectiveness. 
Section \ref{sec:summary} is devoted to the summary. 

\section{Theoretical bounds of Josephson critical currents}
\label{sec:jc}

\begin{figure}[tbp]
\centering
\scalebox{0.40}[0.40]{\includegraphics{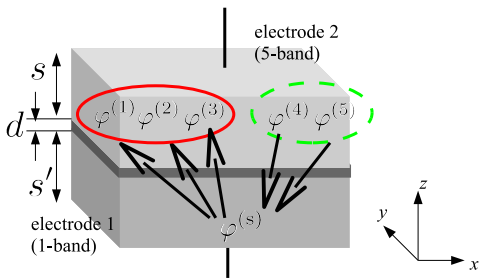}}
\caption{(color online) Schematic diagram of a SIS Josephson junction between
 five-band and single-band superconductors. Here the electrode $2$ is assumed to
 be iron-based superconductor. The three superconducting phases
 enclosed by the solid line can be assigned as hole bands of
 iron-based superconductor, while the two phases enclosed by the dashed
 line as electron bands. }
\label{fig:junction}
\end{figure} 

We examine a superconductor$\mbox{-}$insulator$\mbox{-}$superconductor (SIS) Josephson junction, as
shown in Fig.\,\ref{fig:junction}.  
The electrode $1$ ($2$), whose length is $s^{\prime}$ ($s$) in the
direction of the $z$ axis, is a single-band (five-band) superconductor. 
The insulator, whose length is $d$ in the direction of the $z$ axis and
the dielectric constant is $\epsilon$, is sandwiched between the two
different superconducting electrodes. 

\begin{figure}[bp]
\centering
\scalebox{0.40}[0.40]{\includegraphics{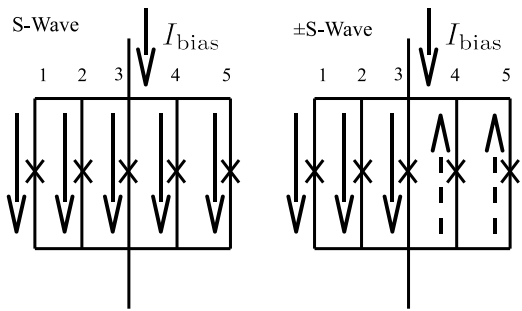}}
\caption{Electric circuits corresponding to a SIS Josephson
 junction. The circuit has the five parallel branches associated with the tunneling
 channels. The bias current is denoted as $I_{\rm bias}$. 
The left panel corresponds to the case of the $s$-wave, while the right
 panel to the $\pm s$-wave. }
\label{fig:circuit}
\end{figure} 

\begin{table*}[tbp]
\begin{tabular}{cccccc}
\hline\hline
 & $N_{1}/N_{\rm tot}$& 
$N_{2}/N_{\rm tot}$& 
$N_{3}/N_{\rm tot}$& 
$N_{4}/N_{\rm tot}$& 
$N_{5}/N_{\rm tot}$\\
\hline
$\mbox{Ba}\mbox{Fe}_{2}\mbox{As}_{2}$ 
& $0.1687$& $0.2757$& $0.2104$& $0.1803$& $0.1649$\\
$\mbox{LaFeAsO}$
& $0.1133$& $0.3009$& $0.2323$& $0.1608$& $0.1928$\\
\hline\hline
\end{tabular}
\caption{DOS ratios on the Fermi surfaces of iron-based
 materials evaluated by a first-principles
 calculation\,\cite{Nakai;Machida:2009}. 
The first three DOS's ($N_{1}$, $N_{2}$, and $N_{3}$) corresponds to the hole
 bands, while the remaining DOS's ($N_{4}$ and $N_{5}$) to the electric
 bands. } 
\label{tab:dos}
\end{table*}

One of the fundamental quantities characterizing Josephson junctions is
the Josephson critical current density. 
One can find various discussion for the cases including two-band superconductors in
several
references\,\cite{Brinkman;Andersen:2002,Agterberg;Janko:2002,Ota;Matsumoto:2009,Sperstad;Sudbo:2009,Ota;Koyama:2009}. 
However, there is no work on arbitrary $N$-band superconductors. 
This paper treats such a general case. 
Throughout this paper, we assume that there is no relative
superconducting phase fluctuation in multi-band superconducting electrodes. 
In this case, the system is described by an electric circuit as shown
in Fig.\,\ref{fig:circuit}. 
We remark that such a rigid parallel circuit modeling is a good
description as far as the relative phase fluctuations are fully pinned.   
We will discuss corrections of relative phase fluctuations to Josephson
effects elsewhere.  
Under this assumption, the total Josephson critical current density is
given by 
\begin{equation}
 j_{\rm c}=
\sum_{i}j_{i}\cos\chi^{(i1)}_{0}. 
\label{eq:jc}
\end{equation} 
$\chi^{(ij)}_{0}$ is a constant phase between the $i$th and the $j$th
superconducting gaps, which reflects the symmetry of a
static gap solution. 
Equation (\ref{eq:jc}) includes $\pi$ channels when
sign changes occur between the superconducting gaps.  
The basic formalism to derive Eq.\,(\ref{eq:jc}) is shown in
Appendix\,\ref{appdx:phase_eq}. 
For the $s$-wave case, $\chi^{(i1)}_{0}=0$ for any $i$. 
On the other hand, a part of 
$\{\chi^{(i1)}_{0}\}$ should be $\pi$ for the $\pm s$-wave symmetry. 
As an example for the $\pm s$-wave case in the electrode $2$, we take 
\(
\chi^{(21)}_{0}=\chi^{(31)}_{0}=\chi^{(32)}_{0}=0
\) and 
\(
\chi^{(41)}_{0}=\chi^{(51)}_{0}=\pi
\) as schematically shown in Fig.\,\ref{fig:circuit}.   
Equation (\ref{eq:jc}) indicates that $j_{\rm c}$ for the 
$\pm s$-wave symmetry is always smaller than that for the $s$-wave, i.e., 
\(
j_{\rm c}(s\mbox{-wave}) > j_{\rm c}(\pm s\mbox{-wave})
\)\,\cite{Ota;Matsumoto:2009,Ota;Koyama:2009}. 

\begin{table*}[htbp]
\begin{tabular}{cccccccc}
\hline\hline
$(T=0\,\mbox{K})$ & 
$\Delta^{(1)}/\Delta_{\rm max}$& 
$\Delta^{(2)}/\Delta_{\rm max}$& 
$\Delta^{(3)}/\Delta_{\rm max}$& 
$\Delta^{(4)}/\Delta_{\rm max}$& 
$\Delta^{(5)}/\Delta_{\rm max}$&
$2\Delta_{\rm max}/k_{\rm B}T^{\rm iron}_{\rm c}$& 
$\Delta^{({\rm s})}/\Delta_{\rm max}$
\\
\hline
$(\mbox{Ba},\mbox{K})\mbox{Fe}_{2}\mbox{As}_{2}$ 
& $0.9395$& $1$& $0.5189$& $0.9415$& $0.9691$& 
$3.785$& $0.192$\\
$\mbox{LaFeAs}(\mbox{O},\mbox{F})$
& $0.5052$& $1$& $0.2677$& $0.5084$& $0.5300$& 
$4.426$& $0.232$\\
\hline\hline
\end{tabular}
\caption{Superconducting gap amplitude ratios of iron-pnictide materials at
 zero temperature estimated by a five-band quasi-classical theory\,\cite{Nakai;Machida:2009}.} 
\label{tab:gap}
\end{table*}

Let us turn to a microscopic formula for $j_{i}$ in Eq.\,(\ref{eq:jc}). 
First, we give notations. 
As for the electrode $1$, we denote the DOS on the
Fermi surface and the superconducting gap amplitude as,
respectively, $N_{\rm s}$ and $\Delta^{({\rm s})}(>0)$. 
Similarly, as for the electrode $2$, $N_{i}$ and $\Delta^{(i)}(>0)$ are
the $i$th DOS and the $i$th gap amplitude, respectively. 
In addition, we define ``smaller'' and ``larger'' gaps as 
\(
\Delta_{{\rm S},i}
=\min\{\Delta^{(i)},\Delta^{(\rm s)}\}
\) and 
\(
\Delta_{{\rm L},i}
=\max\{\Delta^{(i)},\Delta^{(\rm s)}\}
\), 
respectively. 
Thus, assuming the full gap solutions in both superconducting electrodes, we
microscopically calculate $j_{i}$ using a standard second order
perturbation theory with respect to a tunneling
channel\,\cite{Ambegaokar;Baratoff:1963}. 
Then, we have 
\begin{equation*}
j_{i}
=
\frac{1}{W}
\frac{1}{r_{{\rm n},i}}
\frac{\pi\Delta_{{\rm eff},i}}{2e},
\end{equation*}
where 
\begin{equation*}
\frac{1}{r_{{\rm n},i}}
=
\frac{4\pi e^{2}}{\hbar}|T^{(i)}|^{2}N_{\rm s}N_{i}, 
\quad
\Delta_{{\rm eff},i} 
=
\frac{2}{\pi}K(k_{i};\beta\Delta_{{\rm L},i})
\Delta_{{\rm S},i}, 
\end{equation*} 
and $W$ is the area of the junction interface.  
In the definition of $r_{{\rm n},i}^{-1}$, the tunneling 
constant associated with the $i$th tunneling channel is denoted as $T^{(i)}$. 
The quantity $k_{i}$ corresponds to the ratio of the smaller gap to the
larger one,  
\(
k_{i}=[1-(\Delta_{{\rm S},i}/\Delta_{{\rm L},i})^{2}]^{1/2}
\). 
The function $K(k;\nu)$ is given by 
\begin{equation*}
K(k;\nu)
=
\int_{0}^{1}
\frac{\tanh(\nu\sqrt{1-k^{2}x^{2}}/2)}
{\sqrt{(1-k^{2}x^{2})(1-x^{2})}} \,dx.
\end{equation*}

Combining the above arguments with Eq.\,(\ref{eq:jc}), we obtain 
\begin{equation}
 I_{\rm c}R_{\rm n}
=
\sum_{i}
\frac{R_{\rm n}}{r_{{\rm n},i}}\frac{\pi\Delta_{{\rm eff},i}}{2e}
\cos\chi^{(i1)}_{0},
\label{eq:jcrn}
\end{equation}
where $I_{\rm c}=j_{\rm c}W$.  
The combined resistance $R_{\rm n}=1/\sum_{i}r_{{\rm n},i}^{-1}$ can be
experimentally measured when a bias current is greater than 
$I_{\rm c}$, while the individual measurement of $r_{{\rm n},i}$ is
practically impossible. 
Equation (\ref{eq:jcrn}) is a generalized formula of the
Ambegaokar-Baratoff relation for multi-channel heterojunctions.  
Brinkman {\it et al.}\,\cite{Brinkman;Andersen:2002} and 
Agterberg {\it et al.}\,\cite{Agterberg;Janko:2002} obtained similar
results in the context of $\mbox{MgB}_{2}$, i.e., two-band
superconductor. 

\begin{figure}[htbp]
\centering
\scalebox{0.73}[0.73]{\includegraphics{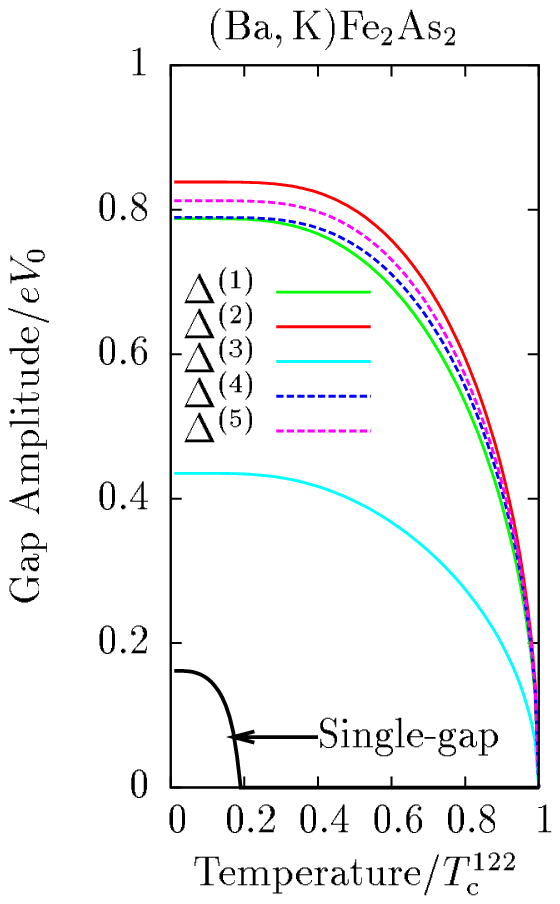}}\hspace{2mm}
\scalebox{0.73}[0.73]{\includegraphics{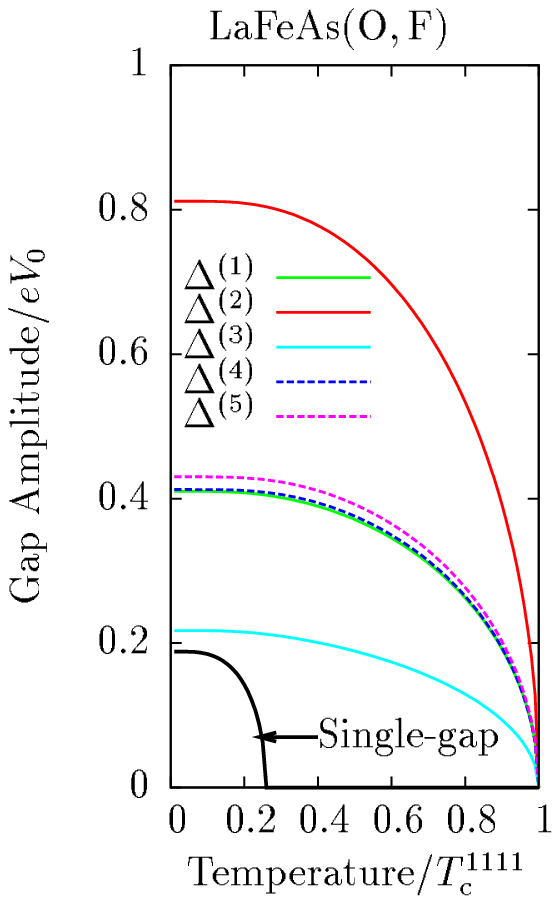}}
\vspace{-4mm}
\begin{flushleft}
(a)\hspace{40mm}(b)
\end{flushleft}
\caption{(color online) Temperature dependence of iron-pnictide and single-band 
 superconductor gap amplitudes. As for the iron-pnictide materials, we
 employ the previous results\,\cite{Nakai;Machida:2009}. The gap
 amplitudes are normalized by  
$eV_{0}=e[\Delta^{({\rm s})}(0)+\Delta_{\rm max}(0)]$. The solid lines
 ($\Delta^{(1)}$, $\Delta^{(2)}$, and $\Delta^{(3)}$) correspond to the
 hole band, while the dashed lines ($\Delta^{(4)}$ and $\Delta^{(5)}$)
 the electric band. In the case of the $\pm s$-wave symmetry, the gaps
 corresponding to the dashed lines have the relative minus signs (i.e.,
 $\chi_{0}^{(41)}=\chi_{0}^{(51)}=\pi$). 
(a) $(\mbox{Ba},\mbox{K})\mbox{Fe}_{2}\mbox{As}_{2}$ 
($T^{122}_{\rm c}=38~\mbox{K}$) and 
(b) $\mbox{LaFeAs}(\mbox{O},\mbox{F})$ 
($T^{1111}_{\rm c}=27~\mbox{K}$). } 
\label{fig:tempdep_gap}
\end{figure}
\begin{figure}[bp]
\centering
\scalebox{0.75}[0.75]{\includegraphics{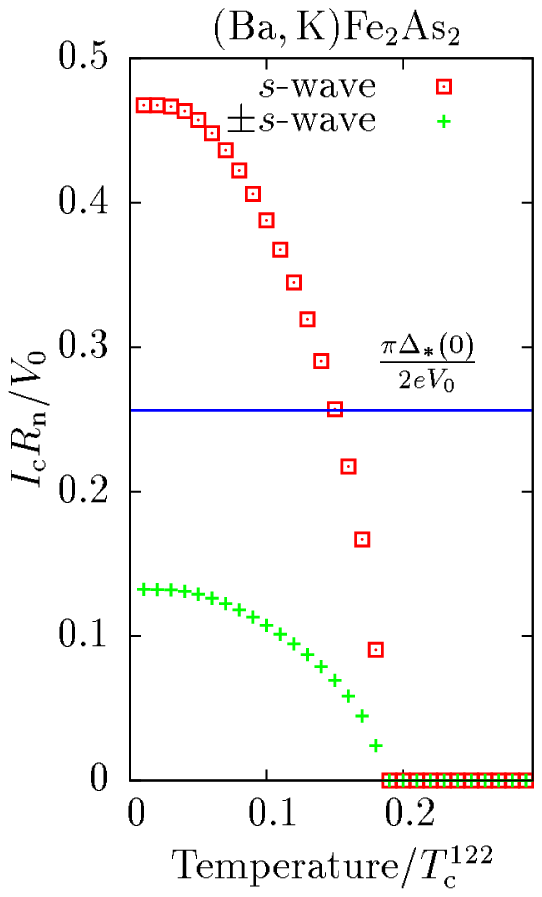}}\hspace{2mm}
\scalebox{0.75}[0.75]{\includegraphics{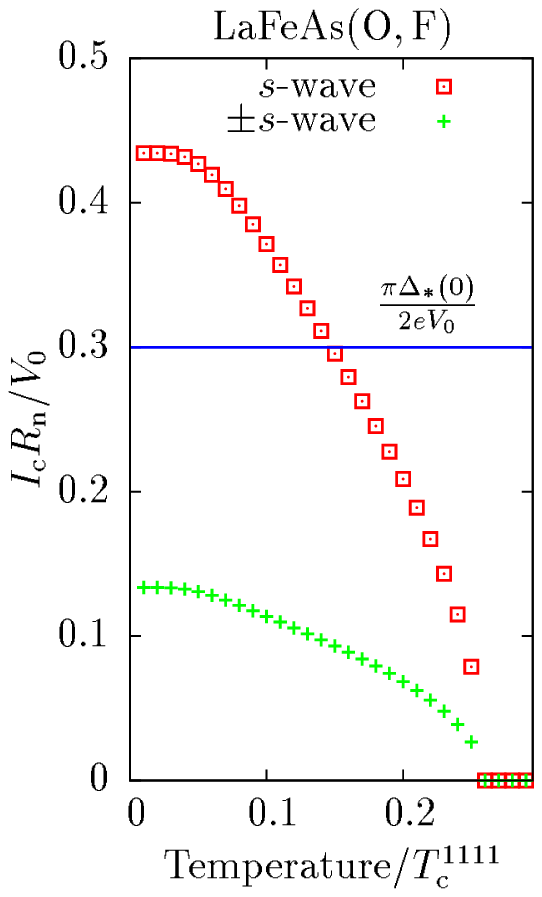}}
\vspace{-7mm}
\begin{flushleft}
(a)\hspace{40mm}(b)
\end{flushleft}
\caption{(color online) $I_{\rm c}R_{\rm n}$ products normalized by 
Eq.\,(\ref{eq:def_vzero}). The red squares are for the
 $s$-wave, while the green crosses are for
 the $\pm s$-wave symmetry (i.e., $\chi_{0}^{(21)}=\chi_{0}^{(31)}=0$ and
 $\chi_{0}^{(41)}=\chi_{0}^{(51)}=\pi$). We also show the theoretical
 lower bounds for the $s$-wave symmetry at zero temperature
 ($\pi\Delta_{\ast}(0)/2eV_{0}$), depicted as the blue solid lines. 
We find that $I_{\rm c}R_{\rm n}/V_{0}$ for $\pm s$-wave is much smaller
 than the theoretical lower bound for the $s$-wave at zero temperature. 
(a) $(\mbox{Ba},\mbox{K})\mbox{Fe}_{2}\mbox{As}_{2}$ 
($T^{122}_{\rm c}=38~\mbox{K}$) and 
(b) $\mbox{LaFeAs}(\mbox{O},\mbox{F})$ 
($T^{1111}_{\rm c}=27~\mbox{K}$). } 
\label{fig:icrn}
\end{figure}
\begin{figure*}[tbp]
\centering
\scalebox{0.73}[0.73]{\includegraphics{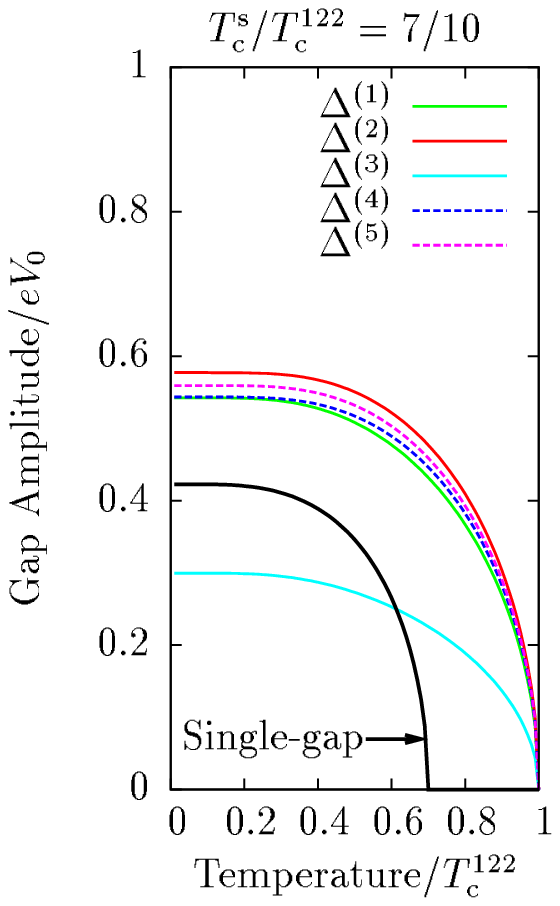}}\hspace{2mm}
\scalebox{0.73}[0.73]{\includegraphics{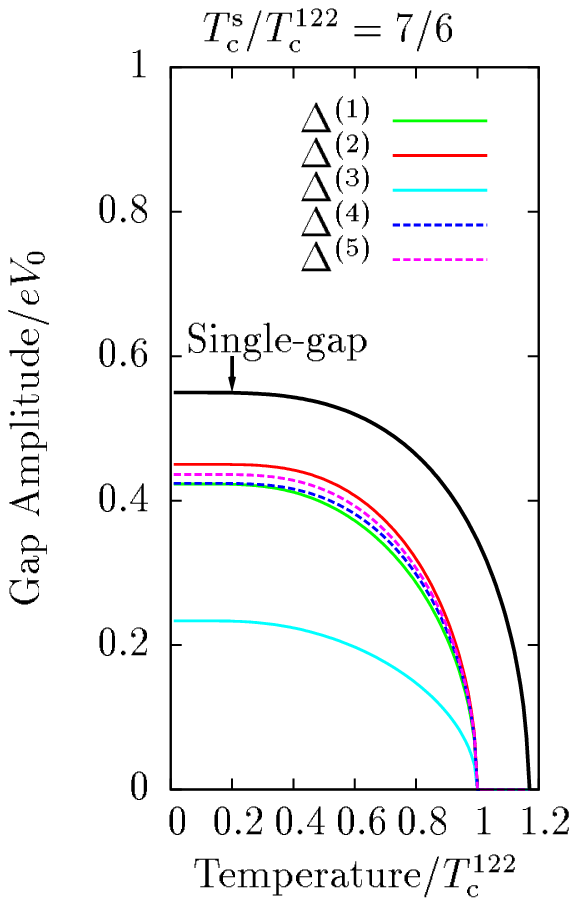}}\hspace{2mm}
\scalebox{0.73}[0.73]{\includegraphics{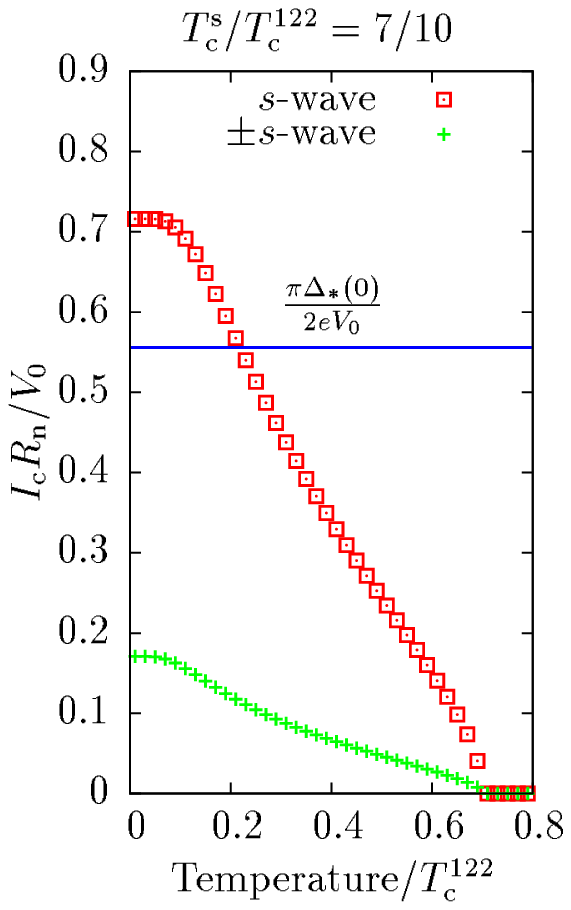}}\hspace{2mm}
\scalebox{0.73}[0.73]{\includegraphics{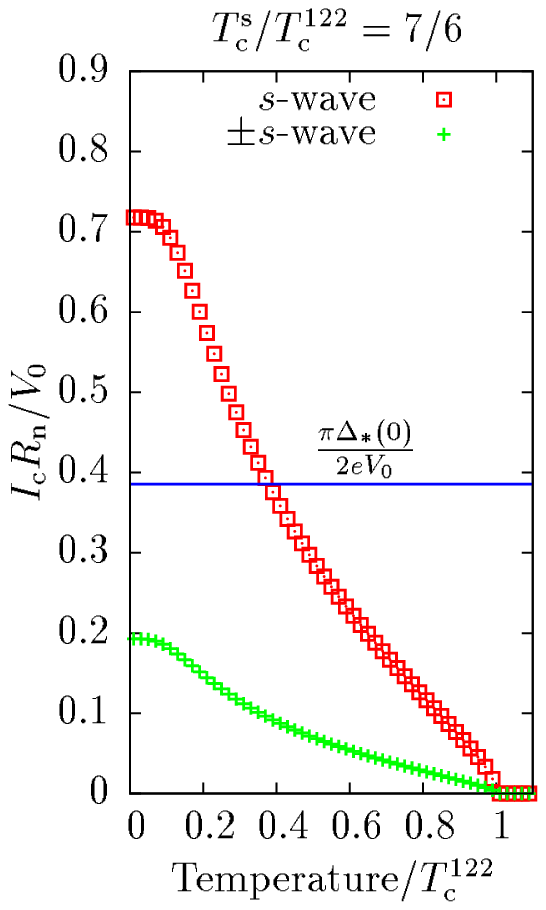}}
\vspace{-7mm}
\begin{flushleft}
(a)\hspace{40mm}(b)
\hspace{40mm}(c)
\hspace{40mm}(d)
\end{flushleft}
\caption{(color online) Results for smaller iron-pnictide superconducting gap
 amplitudes. The DOS and the gap ratios are the same as the ones in
 Fig.\,\ref{fig:icrn}(a), but the iron-pnictide superconducting transition
 temperatures are much smaller than in the previous example. 
The $I_{\rm c}R_{\rm n}$ products normalized by
 Eq.\,(\ref{eq:def_vzero}) are shown in (c) ((d)), in which the
 gap amplitudes drawn in (a) ((b)) are employed.}  
\label{fig:lowTc}
\end{figure*}
\begin{figure}[bp]
\centering
\scalebox{0.8}[0.8]{\includegraphics{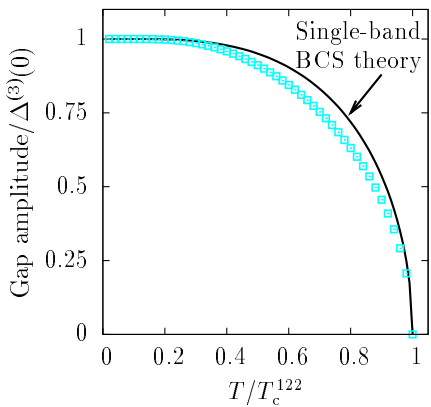}}
\caption{(color online) Comparison of iron-pnictide's gap function to the
 single-band BCS gap function. The cyan square is the small gap $\Delta^{(3)}$
 calculated by the five-band quasi-classical theory for 
$\mbox{(Ba,K)}\mbox{Fe}_{2}\mbox{As}_{2}$\,\cite{Nakai;Machida:2009},
 while the black solid line corresponds to 
 Eq.\,(\ref{eq:singleBCS}), in which we use $T^{122}_{\rm c}$ instead of
 $T^{\rm s}_{\rm c}$. One finds that a difference between the two
 functions exists around $T/T^{122}_{\rm c}=0.4$-$0.8$. We can find a
 similar discrepancy for $\Delta^{(1)}$, which is much
 smaller than in the present figure. }
\label{fig:compare}
\end{figure}

Here, we derive a simple relation from Eq.\,(\ref{eq:jcrn}) when the electrode
$2$ has the $s$-wave symmetry. 
The resultant expression can give useful information about the pairing
symmetry.  
The right hand side of Eq.\,(\ref{eq:jcrn}) is a summation of positive
quantities in this case (i.e., $\chi^{(i1)}_{0}=0$ for any $i$). 
Accordingly, we find that 
\begin{equation}
 I_{\rm c}(s\mbox{-wave})R_{\rm n}
\ge
\frac{\pi \Delta_{\ast}}{2e},
\quad
\Delta_{\ast}=\min_{i} \Delta_{{\rm eff},i},
\label{eq:lowerb_pps}
\end{equation}
which gives a theoretical lower bound of $I_{\rm c}R_{\rm n}$
for the $s$-wave. 
From the above argument, we find that if the symmetry of the electrode
$2$ is $s$-wave then Eq.\,(\ref{eq:lowerb_pps})
must be fulfilled. 
Namely, if a measured value of $I_{\rm c}R_{\rm n}$ satisfies 
the inequality 
\begin{equation}
 I_{\rm c}R_{\rm n}
<
\frac{\pi \Delta_{\ast}}{2e}, 
\label{eq:upperb_pms}
\end{equation}
then one can exclude the possibility of the $s$-wave. 
This method does not give more detailed information about the symmetry
but clarify whether $\pi$ channels exist in the multiple tunneling
junctions with iron-pnictides.  
The present scheme, ``lower-bound criterion'' is a simple
and convenient classification of iron-based superconducting materials. 

Let us discuss experimental applicability of the criterion. 
The direct experimental data for bulk superconduting samples, e.g.,
angle resolved photoemission
spectroscopy\,\cite{Ding;Wnag:2008,Evtushinsky;Borisenko:2009}, provides
the gap amplitudes in the 1-gap and the iron-based superconducting electrodes. 
Therefore, in principle, one can input $\Delta_{\ast}$ in
Eq.\,(\ref{eq:lowerb_pps}).  
However, superconduting gap amplitudes relevant to Josephson
junctions is generally smaller than the ones in the bulk superconductors due
to the damage piled up in interface fabrication. 
It indicates that direct comparison of a measured $I_{\rm c}R_{\rm n}$
product to $\Delta_{\ast}$ based on the bulk data may give no practical information. 
Therefore, we focus on a transition voltage of the all tunneling channels
into running state at zero temperature\,\cite{Schmidt:1997}, 
\begin{equation}
 V_{0}=\frac{1}{e}[\Delta^{({\rm s})}(T=0)+\Delta_{\rm max}(T=0)], 
\label{eq:def_vzero}
\end{equation}
in which 
\(
\Delta_{\rm \max}=\max_{i}\Delta^{(i)}
\). 
One can measure the suppressed $V_{0}$ experimentally if the
junction shows hysteretic $I$-$V$ characteristics\,\cite{Schmidt:1997}. 
We emphasize that a scaled quantity $\Delta_{\ast}/eV_{0}$ is given by
only the gap amplitude ratios since the bulk gap ratios are
expected to be kept as long as the damage is not too severe. 
Therefore, the scaled lower bound $\Delta_{\ast}/eV_{0}$ is
experimentally evaluated by the gap amplitude ratios for the
bulk samples, and this quantity should be compared to measured 
$I_{\rm  c}R_{\rm n}/V_{0}$. 
Consequently, we conclude that the inequality
(\ref{eq:upperb_pms}) is an effective formula to examine the pairing
symmetry. 

\section{Application of lower-bound criterion to Iron-pnictide superconductors}
\label{sec:iron}
Let us evaluate the right hand side of Eq.\,(\ref{eq:jcrn}) in 
real iron-pnictide materials and check how the criterion based on
the inequality (\ref{eq:upperb_pms}) works. 
For the sake of simplicity, we employ a simple model for
the tunneling constants.  
Namely, we assume that $T^{(i)}$'s take channel-independent constants. 
It means that $R_{\rm n}/r_{{\rm n},i}$ is equal to 
$N_{i}/N_{\rm tot}$, where $N_{\rm tot}=\sum_{i}N_{i}$. 
Then, we find that 
$I_{\rm c}R_{\rm n}/V_{0}$ is simply a function of the superconducting
gap ratios (i.e., 
$\Delta^{({\rm s})}/\Delta_{\rm max}$ 
and 
$\Delta^{(i)}/\Delta_{\rm max}$) and the DOS ratios 
(i.e., $N_{i}/N_{\rm tot}$). 

We concentrate on two iron-pnictide superconducting
materials, 
$(\mbox{Ba},\mbox{K})\mbox{Fe}_{2}\mbox{As}_{2}$ 
($T^{122}_{\rm c}=38\,\mbox{K}$) and 
$\mbox{LaFeAs}(\mbox{O},\mbox{F})$ 
($T^{1111}_{\rm c}=27\,\mbox{K}$). 
Here, we denote each superconducting transition temperature
as $T^{122}_{\rm c}$ or $T^{1111}_{\rm c}$.  
As for the single-gap superconducting electrode material, we choose an alloy
of $\mbox{Pb}$, i.e., $\mbox{Pb-In-Au}$. 
The superconducting transition temperature 
\(
T^{\rm s}_{\rm c}=7\,\mbox{K}
\) and 
\(
\Delta^{({\rm s})}(0)/k_{\rm B}T^{\rm s}_{\rm c}
=
1.98
\), respectively\,\cite{Hinken:1991}. 
The gap amplitude ratios and their temperature dependence of
$(\mbox{Ba},\mbox{K})\mbox{Fe}_{2}\mbox{As}_{2}$ and 
$\mbox{LaFeAs}(\mbox{O},\mbox{F})$ 
were evaluated by five-band quasi-classical theory combined with
the DOS's via the first-principles calculations and several kinds of
experimental data for the bulk properties\,\cite{Nakai;Machida:2009}. 
We label three hole bands as $1$ to
$3$ and two electron ones as $4$ to $5$.   
The sign changes are assumed to occur between the hole and the electron bands. 
The DOS and the gap ratios are summarized in Tables \ref{tab:dos} and
\ref{tab:gap}, respectively.  
As for the temperature dependence of the single-band superconducting gap,
we utilize the BCS type gap formula\,\cite{Carrington;Manzano:2003}
\begin{equation}
\Delta^{({\rm s})}(T) 
=
\Delta^{({\rm s})}(0) 
\tanh
\left\{
A
\left[
B
\left(
\frac{T^{\rm s}_{{\rm c}}}{T}-1
\right)
\right]^{C}
\right\},
\label{eq:singleBCS}
\end{equation}
where $A=1.82$, $B=1.018$, and $C=0.51$. 

Figure \ref{fig:tempdep_gap} displays the temperature dependence of the
superconducting gap amplitudes. 
Figure \ref{fig:icrn} shows the temperature dependence of 
$I_{\rm c}R_{\rm n}$ normalized by $V_{0}$. 
The lower bound at $T=0$ given by the right hand side of the inequality
(\ref{eq:lowerb_pps}) is also depicted by the (blue) horizontal line.  
Due to 
$j_{\rm c}(s\mbox{-wave})> j_{\rm c}(\pm s\mbox{-wave})$, 
$I_{\rm c}R_{\rm n}$ for the $\pm s$-wave becomes 
smaller than the one for the $s$-wave over all temperature regions. 
Here, let us focus on the zero temperature. 
$I_{\rm c}R_{\rm n}/V_{0}$ for the $\pm s$-wave is smaller than the
lower bound ($\pi\Delta_{\ast}(0)/2eV_{0}$) while that for $s$-wave is
larger for both of the iron-pnictide superconducting materials. 
Hence, the comparison of a measured value of $I_{\rm c}R_{\rm n}$ to the
theoretical bound gives a useful criterion for the symmetry
in the iron-based superconductors. 

Next, we investigate specific cases in which a part of or all
iron-based superconducting gaps are smaller than the single-band BCS
gap, as shown in Fig.\,\ref{fig:lowTc}(a) and \ref{fig:lowTc}(b). 
In such a study, we keep the gap ratios to be the
values for $\mbox{(Ba,K)Fe}_{2}\mbox{As}_{2}$ shown in Tables
\ref{tab:dos} and \ref{tab:gap} except for 
$\Delta^{({\rm s})}/\Delta_{\rm max}$. 
Figures \ref{fig:lowTc}(c) and \ref{fig:lowTc}(d) show the 
$I_{\rm c}R_{\rm n}$'s for $T^{122}_{\rm c}=10~\mbox{K}$ and
$6~\mbox{K}$, respectively. 
We find again that the inequality (\ref{eq:upperb_pms}) is fulfilled at
zero temperature. 
It means that the present criterion works in every case. 
On the other hand, we notice that the temperature dependences of 
$I_{\rm c}R_{\rm n}$ are relatively anomalous.  
We find a convex behavior in the middle temperature range (i.e.,
$T/T^{122}_{\rm c}\sim 0.2$-$0.8$) contrary to our naive expectation. 
Such peculiarity comes from a discrepancy between the gap
functions calculated by the five-band quasi-classical theory and the 
BCS type gap formula (\ref{eq:singleBCS}), as
shown in Fig.\,\ref{fig:compare}. 
In the previous case as Fig.\,\ref{fig:icrn}, the single-band
BCS gap solely provides the contributions to the temperature dependence
of $I_{\rm c}R_{\rm n}$. 
However, the present case reflects all the gap temperature
dependences. 
Thus, the temperature dependence of $I_{\rm c}R_{\rm n}$ is found to be quite
sensitive to the difference of the gap functions from the single-band
BCS type gap formula\,\cite{commentPRB80e144507}.  

\section{Summary}
\label{sec:summary}
We derived the Ambegaokar-Baratoff relation in the SIS Josephson
junction with multiple (more than two) tunneling channels and proposed a
criterion to identify the pairing symmetry of the iron-based 
superconductors. 
If a measured value of $I_{\rm c}R_{\rm n}$
product is smaller than the lower bound for the $s$-wave, then one
concludes that the symmetry of the 
superconducting electrode is the $\pm s$-wave. 
We actually revealed that the criterion well works in the typical 
iron-pnictide superconductors by employing the DOS and the gap ratios
calculated by the five-band quasi-classical theory and the
first-principles calculation. 
In addition, the theory predicted that the temperature
dependence of $I_{\rm c}R_{\rm n}$ is sensitive to the deviation of the
temperature dependence of the gap from
the single-band BCS formula. 
The method is simple and convenient in contrast to the phase sensitive
measurement like $\pi$ junctions which require much more elaborate setups. 

The Ambegaokar-Baratoff relation is one of the fundamental identity in
the SIS Josephson junction. 
The present analysis suggested that this type of the relation provides
more fruitful information when the junction has multiple tunneling
channels. 

\begin{acknowledgments}
The authors (YO and MM) wish to acknowledge valuable discussion with
S. Shamoto, N. Hayashi, Y. Nagai, M. Okumura, and R. Igarashi. 
They also would like to thank I. Kakeya for helpful comments.  
The work was partially supported by Grant-in-Aid for Scientific Research
on Priority Area ``Physics of new quantum phases in superclean
materials'' (Grant No. 20029019) from the Ministry of Education,
Culture, Sports, Science and Technology of Japan. 
\end{acknowledgments}

\appendix
\section{Formalism for gauge-invariant phase differences}
\label{appdx:phase_eq}
A basic formalism for the hetero Josephson junction dealt in this
paper is presented. 
Assuming uniformity along $y$ axis, the effective Lagrangian density
on the $zx$
plane\,\cite{Ota;Matsumoto:2009}
is given by 
\begin{eqnarray}
&&
 \mathcal{L}_{\rm eff}
=
\frac{s^{\prime}}{8\pi \mu^{\prime\,2}}(q^{0})^{2}
+
\sum_{i=1}^{5}\frac{s}{8\pi \mu_{i}^{2}}
(q^{0}_{i} )^{2} 
+
\sum_{i=1}^{5}\frac{\hbar j_{i}}{e^{\ast}}\cos\theta^{(i)} 
\nonumber \\
&&
\qquad\qquad
+
\sum_{i<i^{\prime}}\frac{\hbar J_{ii^{\prime}}}{e^{\ast}}\cos\chi^{(i^{\prime}i)}
+
\frac{d\epsilon}{8\pi} (E_{21}^{z})^{2},
\label{eq:eff_lag}
\end{eqnarray}
where 
\(
\theta^{(i)} 
= 
\varphi^{(i)} - \varphi^{(\rm s)} 
(e^{\ast}d/\hbar c)A_{21}^{z}
\) and 
\(
\chi^{(i^{\prime}i)}
=
\varphi^{(i^{\prime})}-\varphi^{(i)}
=
\theta^{(i^{\prime})}-\theta^{(i)}
\). 
$\chi^{(i^{\prime}i)}$ is the relative phase difference between the
different superconducting gaps. 
The first and the second terms in Eq.\,(\ref{eq:eff_lag}) represent
charge compressibility in the
electrode $1$ and $2$, respectively, where   
\(
q^{0} = (\hbar/e^{\ast})\pdt\varphi^{(\rm s)} + A_{1}^{0}
\), 
\(
a_{i}^{0} = (\hbar/e^{\ast})\pdt\varphi^{(i)} + A_{2}^{0}
\), and $e^{\ast}=2e$. 
The electric scalar potential in the electrode $\ell(=1,2)$ is denoted
as $A_{\ell}^{0}$, and the charge screening length in the electrode $1$
($2$) is written as $\mu^{\prime}$ ($\mu_{i}$). 
The third term in Eq.\,(\ref{eq:eff_lag}) is the Josephson coupling, in
which $\hbar j_{i}/e^{\ast}$ is the coupling constant associated with
the $i$th tunneling channel. 
The forth term in Eq.\,(\ref{eq:eff_lag}) is called the inter-band Josephson
coupling energy\,\cite{Leggett:1966}, whose origin is the inter-band interaction
between different bands in the electrode $2$. 
In the gauge-invariant phase difference $\theta^{(i)}$, 
the vector potential in the insulator 
\(
A_{21}^{z} = d^{-1}\int_{-d/2}^{d/2}A^{z}(z) dz
\), 
in which $A^{z}(z)$ is the $z$-component of the vector potential. 
The electric field is expressed as 
\(
 E_{21}^{z} 
= -c^{-1}\pdt A_{21}^{z} - d^{-1}(A^{0}_{2}-A^{0}_{1})
\). 

As the Euler-Lagrangian equation with respect to
$A^{0}_{\ell}$, we obtain the Josephson relation 
\begin{equation}
 \sum_{i}\frac{\bar{\alpha}}{\alpha_{i}}\pdt\theta^{(i)}
= 
\frac{e^{\ast}\Lambda d}{\hbar} E_{21}^{z},
\label{eq:mod_jr}
\end{equation}
where 
\(
\alpha^{\prime} = \epsilon \mu^{\prime 2}/s^{\prime}d 
\), 
\(
\alpha_{i} = \epsilon \mu_{i}^{2}/sd 
\),  
\(
\bar{\alpha}^{-1} = \sum_{i}\alpha_{i}^{-1}
\) 
and 
$\Lambda=1+\alpha^{\prime}+\bar{\alpha}$. 
Next, as the Euler-Lagrangian equation with respect to
$A^{z}_{21}$, we have the Maxwell equation 
\begin{equation}
0
= 
\frac{\epsilon}{c^{2}}\frac{e^{\ast}d}{\hbar}\pdt E^{z}_{21}
+
\sum_{i}\frac{4\pi e^{\ast}d}{\hbar c^{2}}j_{i}
\sin\theta^{(i)}.
\label{eq:maxwell}
\end{equation}
Combining Eq.(\ref{eq:maxwell}) with Eq.\,(\ref{eq:mod_jr}) leads to the
equation 
\begin{equation}
\frac{\epsilon}{4\pi \Lambda d}\frac{\hbar}{e^{\ast}}
\sum_{i}\frac{\bar{\alpha}}{\alpha_{i}}\pdt^{2}\theta^{(i)}
+
\sum_{i} j_{i}\sin\theta^{(i)}=0.
\label{eq:rcsj}
\end{equation}
If one adds a dissipation and an
external bias current terms to Eq.\,(\ref{eq:rcsj}), one has the
resistively and capacitively shunted 
junction model with
multi-tunneling channels. 

Let us turn to a formula for the Josephson critical current density. 
Generally, the system has a relative superconducting phase
fluctuation originating from the inter-band Josephson coupling 
\(
\sum_{i<i^{\prime}} J_{ii^{\prime}}\cos\chi^{(i^{\prime}i)}
\), which 
generates Josephson-Leggett collective excitation
modes\,\cite{Leggett:1966,Ota;Matsumoto:2009}.  
The equations of motion for $\chi^{(i^{\prime}i)}$ can be obtained as
the Euler-Lagrange equations with respect to 
$\varphi^{(\rm s)}$ and $\varphi^{(i)}$\,\cite{Ota;Matsumoto:2009}. 
In this paper, we simply assume that each $\chi^{(i^{\prime}i)}$ is
fixed as a constant $\chi^{(i^{\prime}i)}_{0}$ reflecting the symmetry
of a static gap solution. 
Such an assumption can be validated when $|J_{ii^{\prime}}| \gg j_{1},j_{2}$. 
In this case, we have 
\(
 \theta^{(i)}(t) = \theta^{(1)}(t) + \chi^{(i1)}_{0}
\), 
where $\chi^{(i1)}_{0}=0$ or $\pi$. 
Equation (\ref{eq:rcsj}) is then rewritten by 
\begin{equation}
\frac{\epsilon}{4\pi \Lambda d}
\frac{\hbar}{e^{\ast}}
\pdt^{2}\theta^{(1)}
+
\sum_{i} j_{i}\cos\chi^{(i1)}_{0}\sin\theta^{(1)}
=j_{\rm bias}. 
\label{eq:reduced_eq}
\end{equation} 
Here, we add a bias current density $j_{\rm bias}$ to the
right hand side. 
Equation (\ref{eq:reduced_eq}) indicates that a static solution exists
as long as $j_{\rm bias}<|\sum_{i}j_{i}\cos\chi^{(i1)}_{0}|$.

\end{document}